\documentclass[aps,prd,twocolumn,groupedaddress,amsmath,amssymb,reprint,10 pt]{revtex4-2}

\usepackage{graphicx}% Include figure files
\usepackage{dcolumn}% Align table columns on the decimal point
\usepackage{bm}% bold math
\usepackage{amsmath}
\usepackage{amssymb}
\usepackage{xcolor}
\usepackage{hyperref}% add hypertext capabilities
\hypersetup{
	colorlinks=true,
	linkcolor=red,
	filecolor=magenta,      
	urlcolor=blue,
	pdftitle={Selective Thermalization},
	pdfpagemode=FullScreen,
	citecolor= blue,
}
\begin{document}

% number in the upper right-hand corner of the title page in preprint mode.
% Use the 'preprint numbers' class option to override journal defaults
%\preprint{}


\title{Generation and purification of excited spacetimes using Schwarzian derivative}

% author. Explanatory text should go in the []'s actual e-mail
% Please use the appropriate macro for each type of information

%\thanks{}
%\affiliation{}
%\author{}
\author{Rakesh K Jha}
	\email{p20230070@hyderabad.bits-pilani.ac.in}
    \author{Akhil U Nair}
    \email{p20200473@hyderabad.bits-pilani.ac.in}
    \author{Prasant Samantray}
    \email{prasant.samantray@hyderabad.bits-pilani.ac.in}
   \author{Sashideep Gutti}
	\email{sashideep@hyderabad.bits-pilani.ac.in}
	\affiliation{Department of Physics, Birla Institute of Technology and Science, Pilani-Hyderabad Campus \\Hyderabad, 500078, India}	
	\date{\today}% It is always \today, today,

%Collaboration name if desired (requires superscriptaddress

\date{\today}

\begin{abstract}
 In this article, we use the expression of the Schwarzian derivative to set up differential equations to find answers to three fundamental questions in the context of QFT in curved spacetime, specifically in two dimensions. One of the ways in which one can derive the Unruh effect in two dimensions is to use the anomalous transformation law of the energy-momentum tensor for a CFT that involves a Schwarzian derivative (Virasoro Anomaly). We answer the following three questions in the context of a massless scalar field in two-dimensional flat spacetime. The first question is as follows: If we have a spacetime with a massless scalar field in vacuum, what are all the subsets of spacetime such that the subset has a thermal distribution of particles for the left-moving and/or right-moving sectors? We obtain a general solution to this question by setting up and solving a third-order nonlinear differential equation based on the expression of Schwarzian. Based on the general solution, we can generate various subsets of the given spacetime that have a thermal flux/density of particles, of which the Rindler spacetime is one. The second question is an inverse question in which we suppose we are given a spacetime with a thermal distribution of particles; what are the possible purifying spacetimes (the ``parent'' spacetimes with the field in vacuum state whose reduced state in the given spacetime yields the observed particle content)? We similarly obtain a general class of solutions by setting up and solving a second differential equation. In this context, we also define ``partial purification'' where we obtain a spacetime that purifies only the left-moving or right-moving sector. The third question concerns locating spacetimes with the same particle content starting from the same ``parent'' spacetime. These sibling spacetimes are generated again by obtaining the general solution of a third differential equation based on the expression of Schwarzian. 
\end{abstract}

%\keywords{}

\maketitle{}

Hawking radiation is a fundamental prediction of quantum field theory in curved spacetime, arising from the presence of horizons and their associated thermodynamic properties~\cite{Hawking1975, BirrellDavies}. A simplified setting for studying these features is provided by Rindler spacetime, which corresponds to uniformly accelerated observers in Minkowski spacetime. Such observers perceive the Minkowski vacuum as a thermal state, giving rise to the Unruh effect~\cite{Unruh:1976db, Birrell:1982ix, Mukhanov:2007zz, Unruh:1983ms}. This illustrates the observer dependence of the particle content and its relation to the horizon structure.

In a two-dimensional conformal field theory, these effects can be analyzed using conformal transformations. The stress-energy tensor transforms anomalously under such maps, with the Schwarzian derivative encoding the Virasoro anomaly~\cite{Fabbri:2005mw, Blumenhagen:2009zz}. This provides a direct framework for relating coordinate transformations to energy flux and particle production.

The close connection between the Schwarzian derivative and the expectation value of the stress-energy tensor suggests the following natural question: Which conformal transformations preserve the energy flux while relating different accelerated observers? Such transformations are of particular interest because they map one thermal description to another without altering the radiative flux measured at the null infinity. Characterizing this class of transformations provides insight into the symmetry structure underlying horizon thermodynamics and observer dependence of quantum radiation.

In recent years, conformal methods based on Virasoro algebra have proven to be powerful tools for analyzing near-horizon physics and the quantum aspects of gravity~\cite{jha2026inequivalentpathsthermalityminkowski}. Rather than treating Hawking or Unruh radiation solely as consequences of quantum field theory in a fixed background, these approaches emphasize the role of asymptotic symmetries~\cite{govindarajan2026quasicharactersthreecharacterrationalconformal} and the anomalous transformation properties of the stress-energy tensor. The Schwarzian derivative~\cite{jha2026inequivalentpathsthermalityminkowski, Fabbri:2005mw, Blumenhagen:2009zz} naturally emerges as the quantity governing the change in energy flux under conformal transformations, making it an ideal framework for studying horizon-induced radiation. Motivated by these ideas, we pose three inverse problems and present the solutions in this article. The analysis is done in two-dimensional Minkowski spacetime with a massless scalar field living in it. \par
The first question is stated broadly as follows: Given a spacetime with a quantum field in vacuum, what are the subsets of this spacetime that have a thermal density of particles?  (like the Rindler spacetime in Minkowski spacetime with a field in a vacuum). We answer this question in the context of two-dimensional Minkowski spacetime with a massless scalar field defined on it. We solve this by formulating an inverse problem. We require that the transformed stress tensor exhibits a constant thermal flux. This yields a third-order nonlinear differential equation based on the expression of Schwarzian. We obtain the general solution for the differential equation. This yields a broader family of Möbius-type transformations~\cite{yan2026formalizingextendedcomplexnumbers} that generate many inequivalent coordinate systems while generating a thermal density of particles (presented in Sec.~\ref{sec-5}). Rindler spacetime is one solution among them.  We further examine the quantum state associated with these transformations and show that the corresponding vacua remain thermally populated with the same radiation flux.  We can also choose the starting point to be a Rindler spacetime, which yields a plethora of interesting subsets of Rindler spacetimes that have a thermal density of particles starting from a Rindler spacetime in vacuum. This is presented in Sec.~\ref{sec-6_1}. \par
The second question we pose is as follows: Given a spacetime with a field in the thermal density of particles, what are the possible parent spacetimes (supersets of this spacetime) whose vacuum state yields the observed particle density? For instance, a Minkowski spacetime with a field in vacuum is the purifying spacetime for a Rindler spacetime with a thermal distribution of particles. In the context of Quantum Field theory, this amounts to the ``purification of mixed states.'' We answer this question in the context of two dimensions. We also introduce the idea of ``partial purification'' in which we purify one of the sectors (left-moving or right-moving sector while leaving the other sector with a thermal flux of particles).  We obtain another third-order nonlinear differential equation based on the expression of Schwarzian. We obtain the general solution, and from it, we can generate various possibilities for the parent spacetimes. Using this solution, we can construct the metric of a purifying spacetime. This is presented in Sec.~\ref{sec-7_1}. \par
The third question is as follows: Given a spacetime with a thermal density of particles, can we locate sibling spacetimes (which share the parent spacetime and have a similar particle content)? For instance, a Rindler spacetime shifted translationally with respect to a given Rindler wedge has the same thermal spectrum of particles when the field is in vacuum in Minkowski spacetime. This problem again yields a third differential equation based on the expression of Schwarzian. We again obtain a general class of sibling spacetimes that can be obtained from a given spacetime with the same particle content. We note that we do not require information regarding the parent spacetime in this scenario. This case is presented in Sec.~\ref{sec-8_1}.

\section{setup\label{sec-2}}
\subsection{Unruh effect and Virasoro anomaly method}

In this section, we recap one of the ways of deriving the Unruh effect using the Virasoro anomaly along the lines presented in~\cite{Fabbri:2005mw}. The starting point of this analysis is the Minkowski spacetime with a massless scalar field in a vacuum state. From the perspective of Rindler spacetime, this vacuum state appears as a thermal flux of particles.  In this section, we introduce the setup and expression for the Schwarzian derivative, which is a crucial ingredient in the calculations presented in this article. The central idea of this study is also discussed in this section.
We consider a two-dimensional Minkowski spacetime with coordinates $T_M, X_M$ with Metric given by $ds^2=-dT_M^2+dX_M^2$. We define the null coordinates given by $U_M=T_M-X_M$ and $V_M=T_M+X_M$. The Metric in these coordinates becomes $ds^2=-dU_MdV_M$. 

We can define a Rindler wedge $R_1$ with coordinates $t_1,x_1$. The transformation between the Rindler-1 coordinates ($R_1$) and Minkowski ($M$) is given by
\begin{equation}
	T_M = \frac{e^{a x_1}}{a} \sinh(a t_1), \label{Eq:2.1.0.1}
\end{equation}
\begin{equation}
	X_M = \frac{e^{a x_1}}{a} \cosh(a_1 t_1). \label{Eq:2.1.0.2}
\end{equation}
We now introduce the light-cone coordinates  in $R_1$, $(u_1,v_1)$:
$
u_1 = t_1 - x_1, \quad v_1 = t_1 + x_1
$.
The lightcone coordinates in Minkowski spacetime are related to the coordinates of $R_1$ as
\begin{equation}
    U_M = T-X = -\frac{e^{-a u_1}}{a}, 
    \label{Eq:2.1.0.3}
\end{equation}
\begin{equation}
    V_M = T+X = \frac{e^{a v_1}}{a}. 
    \label{Eq:2.1.0.4}
\end{equation}
The Metric in Rindler coordinates is
\begin{equation}
    ds^2=-e^{a(v_1-u_1)}du_1dv_1=-e^{2ax_1}(dt_1^2 - dx_1^2).  \label{Eq:2.1.0.5}
\end{equation}
We consider a massless scalar field in vacuum in Minkowski spacetime. The central idea is that the massless scalar field decomposes into the left-moving sector $U_M$ and right-moving sector $V_M$~\cite{Fabbri:2005mw, Jha:2025tpg}. The same split holds for the Rindler spacetime. This is also evident in the map, where the null coordinates $U_M$ maps to $u_1$ and $V_M$ to $v_1$.\\

The transformation law for the energy-momentum tensor is modified in the presence of a central charge in the 2d CFT. The transformation law is as follows: We use the Schwarzian derivative to compute the expectation value of the components of the energy-momentum tensor through the Minkowski Vacuum $|0_M\rangle$, and it is given as
\begin{equation}
  \begin{split}
       \langle 0_{M}|:T_{u_1 u_1}:|0_{M}\rangle  = \bigg(\frac{\partial_{U_M}}{\partial_{u_1}}\bigg)^2  \langle 0_{M}|:T_{U_M U_M}:|0_{M}\rangle\\ - \frac{\hbar}{24\pi} S(U_M,u_1). \label{Eq:2.1.0.6}
  \end{split}
\end{equation}
Where $S(U_M,u_1)$ is the Schwarzian derivative defined by~\cite{Blumenhagen:2009zz, Fabbri:2005mw},
\begin{equation}
    S(U,u) = \frac{1}{ (\frac{\partial U}{\partial u} )^2}\left[(\frac{\partial U}{\partial u})\left(\frac{\partial^3U}{\partial u^3}\right)-\frac{3}{2}\left(\frac{\partial^2 U}{\partial u^2}\right)^2\right]. \label{Eq:2.1.0.7}
\end{equation}
We note that a similar transformation rule is applicable for the left-moving light coordinates $V$ and $v$. More generally, under a conformal transformation that maps from a coordinate chart $(U, V)$ to $(u,v)$ via the relation, $U=U(u)$ and similarly $V=V(v)$, the energy-momentum tensor behaves as
\begin{equation*}
    T^\prime_{uu} = \bigg(\frac{\partial U}{\partial u}\bigg)^2 T_{UU} + \frac{c}{12}  S(U,u),
\end{equation*}
where $c$ denotes the central charge. 
In the context of the Unruh effect, the central charge that works is $c=-\hbar/2\pi$~\cite{Fabbri:2005mw}.

We now recap the steps to justify the Unruh effect~\cite{Fabbri:2005mw} as follows: In the Minkowski Vacuum state $|0_M\rangle$,
\begin{equation}
    \langle 0_M|T_{U_M U_M}|0_M\rangle = 0  \quad\text{\&}\quad  \langle 0_M|T_{V_M V_M}|0_M\rangle = 0 . \label{Eq:2.1.0.8}
\end{equation}
The expectation value of the energy-momentum tensor is defined with respect to the Minkowski Vacuum state $|0_M\rangle$,
\begin{equation}
   \langle 0_M|:T_{u u}:|0_M\rangle = -\frac{\hbar}{24\pi}S(U,u), \label{Eq:2.1.0.9}  
\end{equation}
From  Eqs.~(\ref{Eq:2.1.0.3}),~(\ref{Eq:2.1.0.4}),~(\ref{Eq:2.1.0.6}) and~(\ref{Eq:2.1.0.9}), and after a simple calculation, the expectation value of the energy-momentum tensor in Rindler $\mathrm{R_1}$ gives, 
\begin{equation}
     \langle 0_M|:T_{u_1 u_1}:|0_M\rangle = \frac{\hbar a^2}{48\pi},\label{Eq:2.1.0.10}
\end{equation}
\begin{equation}
    \langle 0_M|:T_{v_1 v_1}:|0_M\rangle = \frac{\hbar a^2}{48\pi}.\label{Eq:2.1.0.11}
\end{equation}
We note that there is a constant flux for both the left- and right-moving sectors.  We also note that the left-moving and right-moving sectors are independent, even though they have the same value for the flux of particles. 
\section{Generation of Rindler-like excited spacetimes from Minkowski spacetime \label{Sec-3}}
As pointed out earlier, the Rindler spacetime is an excellent toy model for analyzing and studying the properties of black hole horizons. Is there a way to discover and generate spacetimes that are similarly excited in terms of particle content, starting from vacuum spacetime?  To this end, we now ask the inverse question: suppose we have a two-dimensional spacetime with null coordinates $(U, V)$. We then ask the question: are there functions $ U (u) $ and $ V (v) $ such that the Schwarzian derivatives $S(U,u)=c_1$ and $S(V,v)=c_2$, where $c_1$ and $c_2$ are constants? We solve this problem using the CFT in the vacuum state in the chart $U, V$.  The strategy is to choose various values of $c_1$ and $c_2$ and determine the subsequent maps. From the maps, we can determine the metric of spacetime and its boundaries.

\subsection{\texorpdfstring{Simplest case: $c_1=0$ and $c_2=0$}{c1=0 and c2=0}\label{subsec-00}}
In this subsection, we solve the Schwarzian differential equation with $c_1=0$ and $c_2=0$. To understand the relevance of this calculation, we considered the following scenario. Suppose that we have a two-dimensional spacetime with a conformal field in the vacuum state. Which other spacetimes register the field in vacuum state? For instance, if we have the CFT in vacuum in Minkowski spacetime, which transformations of the left-moving sector and independently of the right-moving sector produce the same particle content?   Therefore, we solve for $S(Y, X)=0$. The general solution is known as the Möbius  transformation and is given by,
\begin{equation}
    Y=\frac{AX+B}{CX+D},
    \label{Eq:4.1.0.1}
\end{equation}
with the condition that $ad-bc \neq 0$. This is a well-studied scenario, and the transformation above, represented by Eq.~(\ref{Eq:4.1.0.1}), is known as a Möbius transformation \cite{Blumenhagen:2009zz}. 

The Möbius transformation in Eq.~\eqref{Eq:4.1.0.1} has a well-studied physical interpretation. It represents the most general class of coordinate transformations that preserve the vacuum structure of a two-dimensional conformal field theory \cite{Blumenhagen:2009zz}. In particular, these transformations can be thought of as a combination of translations, dilatations, and special conformal transformations, which together form the global conformal group. In Minkowski spacetime, this corresponds to transformations that include Lorentz transformations followed by spacetime translations and scalings, all of which leave the notion of particles invariant \cite{Kim2010Mobius, Christodoulides2014FieldTheoreticalLie,hy9g-pvxv}.

\section{\texorpdfstring{The case $c_1 \neq 0$ and/or $c_2 \neq 0$}{c1 != 0 and/or c2 != 0} \label{sec-5}}
We now solve for the case in which the Schwarzian is constant.  The constants being non-zero implies that there is a non-zero flux of particles in the particular sector (either left-moving, right-moving, or both). We note that this differential equation implies that there is zero particle content in the spacetime with coordinates $ Y$s.
\begin{equation}
(1/ \partial_{X} Y)^2[(\partial_{X} Y)(\partial^3_{X} Y)-\frac{3}{2}(\partial^2_{X} Y)^2]=-a^2.
\label{Eq:5.1.0.1}
\end{equation}
The general solution to the above is given in Appendix~\ref{Apn1}. By choosing various values for $A, B, C, D$,
\begin{equation}
    Y=\frac{Ae^{a X}+B}{{Ce^{a X}+D}}, \label{Eq:5.1.0.2}
\end{equation}
one can obtain all possible spacetimes that have a flux of particles. From the map, we can also construct a metric of spacetime that has a flux of particles.  In the above map, when we disregard cases involving translating the coordinates using constants B or D, we end up with the basic possibilities $Y=e^{ax}/D$ or $Y=e^{-a x}B$. By a suitable choice, we can make $D=\pm a$ and $B=\pm 1/a$. 
\subsection{Maps starting with Minkowski vacuum}
We now choose the parent spacetime to be the Minkowski spacetime with a massless scalar field in vacuum. We use the above map to generate possible subsets of Minkowski spacetimes that have a thermal flux/density of particles. Choosing $Y$  to be $U_M$ ( or $V_M$) and $X$ to be $u_1$ (or $v_1$). We obtain the following conformal maps:
\subsubsection{Half-sided right movers map}
We choose the map $U_m=-e^{-au_1}/a$ and $V_M=v_1$.
\begin{figure}[h!]
    \centering
    \includegraphics[width=0.8\linewidth]{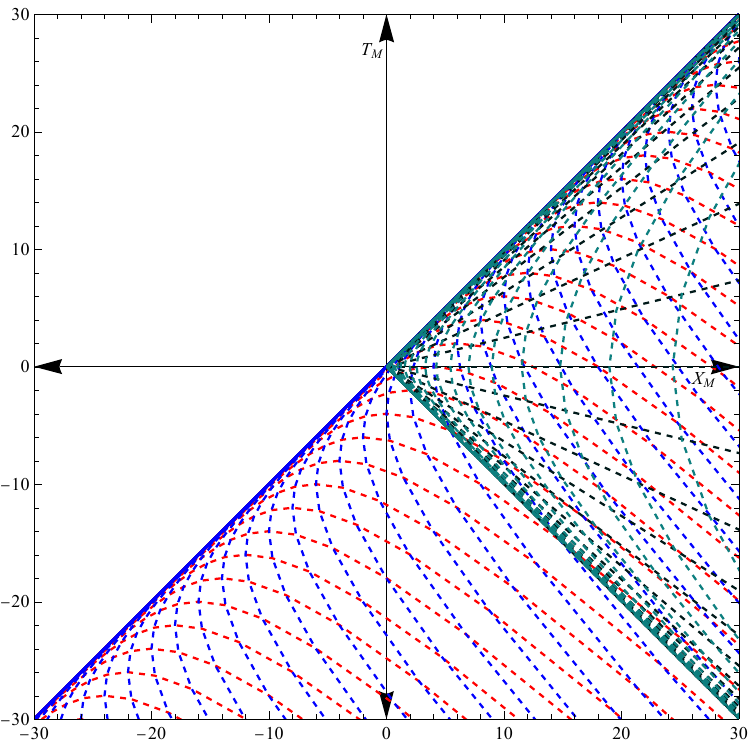}
    \caption{Half-sided right movers map: The spacetime has a thermal flux of right-moving particles}
    \label{Fig_1}
\end{figure}

Given the map into new coordinates is $(u_1,v_1)$, the Metric in these new coordinates can be written as,
\begin{equation*}
    ds^2 = - \bigg(-\frac{1}{a} e^{-a u_1}(-a) du_1\bigg)\big(dv_1\big).
\end{equation*}
After further simplification, the above expression can be written as,
\begin{equation}
    ds^2 = - e^{-a u_1} du_1 dv_1, \label{Eq:5.1.0.3}
\end{equation}
Eq.~(\ref{Eq:5.1.0.3}), can be written in $(t,x)$-coordinate,
\begin{equation}
    ds^2 =  - e^{-a(t_1 - x_1)}\;(dt_1^2 - dx_1^2).  \label{Eq:5.1.0.4}
\end{equation}
The metric given in Eq.~(\ref{Eq:5.1.0.4}) is illustrated in Fig.~\ref{Fig_1}. The right-moving mode in Minkowski coordinates is expressed as,
\begin{equation}
    \hat{\phi}(U_M) = \int_0^\infty \frac{d\omega_0}{\sqrt{4\pi\omega_0}}\bigg(\overrightarrow{\hat{a_0}}(\omega_0) e^{-i\;\omega_0\; U_M}  +\overrightarrow{\hat{a_0}^\dagger}(\omega_0)e^{i\;\omega_0\; U_M} \bigg) ,  \label{Eq:5.1.0.5}
\end{equation}
and the left-mover mode in the Minkowski coordinate is written as, 
\begin{equation}
    \hat{\phi}(V_M) = \int_0^\infty \frac{d\omega_0}{\sqrt{4\pi\omega_0}}\bigg(\overleftarrow{\hat{b_0}}(\omega_0) e^{-i\;\omega_0\; V_M}  +\overleftarrow{\hat{b_0}^\dagger}(\omega_0)e^{i\;\omega_0\; V_M} \bigg).  \label{Eq:5.1.0.6}
\end{equation}
Similarly, the right-mover mode in the Rindler coordinate ($\mathbf{R_1}$) is written as,
\begin{equation}
    \hat{\phi_1}(u_1) = \int_0^\infty \frac{d\omega}{\sqrt{4\pi\omega}}\bigg(\overrightarrow{\hat{a_1}}(\omega) e^{-i\;\omega\; u_1}  +\overrightarrow{\hat{a_1}^\dagger}(\omega)e^{i\;\omega\; u_1} \bigg) ,  \label{Eq:5.1.0.7}
\end{equation}
 and the left-mover mode is: 
 \begin{equation}
    \hat{\phi_1}(v_1) = \int_0^\infty \frac{d\omega}{\sqrt{4\pi\omega}}\bigg(\overrightarrow{\hat{b_1}}(\omega) e^{-i\;\omega\; v_1}  +\overrightarrow{\hat{b_1}^\dagger}(\omega)e^{i\;\omega\; v_1} \bigg).  \label{Eq:5.1.0.8}
\end{equation}
The Bogolyubov coefficients can be obtained by performing the Fourier transform for both the right and left mover modes in Rindler and Minkowski spacetime ( see Appendix~\ref {Apn2}), and after simplification, we obtain,
\begin{equation}
    \alpha^*_{10}(\omega,\omega_0) = \frac{1}{2\;\pi\;a}\sqrt{\frac{\omega}{\omega_0}}\Gamma\bigg[\frac{-i\omega}{a}\bigg]\,e^{\frac{\pi\,\omega}{2a}}\,\bigg(\frac{\omega_0}{a}\bigg)^\frac{i\,\omega}{a} .  \label{Eq:5.1.0.9}
\end{equation}
Similarly,
\begin{equation}
    \beta^*_{10}(\omega,\omega_0) = -\frac{1}{2\;\pi\;a}\sqrt{\frac{\omega}{\omega_0}}\Gamma\bigg[\frac{-i\omega}{a}\bigg] \,e^{-\frac{\pi\,\omega}{2a}}\,\bigg(\frac{\omega_0}{a}\bigg)^\frac{i\,\omega}{a}. \label{Eq:5.1.0.10}
\end{equation}
Now we evaluate the thermal flux. The Minkowski Vacuum is in coordinates $\big(U_M, V_M\big)$, hence the Schwarzian derivative for the map $U_m=-e^{-au_1}/a$ is,
\begin{equation}
    S(U_M,u_1) = -\frac{a^2}{2} .  \label{Eq:5.1.0.11}
\end{equation}
The expectation value of the components of the stress-energy tensor in the Minkowski vacuum is,
\begin{equation}
    \langle 0_M|:T_{U_MU_M}:| 0_M\rangle = \langle 0_M|:T_{V_MV_M}:| 0_M\rangle = 0 .   \label{Eq:5.1.0.12}
\end{equation}
Substituting the above values in Eq.~(\ref{Eq:2.1.0.6}),and after further simplification, we obtain
\begin{equation}
    \langle 0_M|:T_{u_1u_1}:| 0_M\rangle  = \frac{\hbar a^2}{48\pi}.  \label{Eq:5.1.0.13}
\end{equation}
Similarly, the expectation value of the stress energy tensor for the map $V_M=v_1$ is
\begin{equation}
    \langle 0_M|:T_{v_1v_1}:| 0_M\rangle  =  0.   \label{Eq:5.1.0.14}
\end{equation}
This spacetime has a flux of right-moving particles and no left-moving particles. 
\subsubsection{Half-sided  left movers map}

\begin{figure}[h!]
    \centering
    \includegraphics[width=0.8\linewidth]{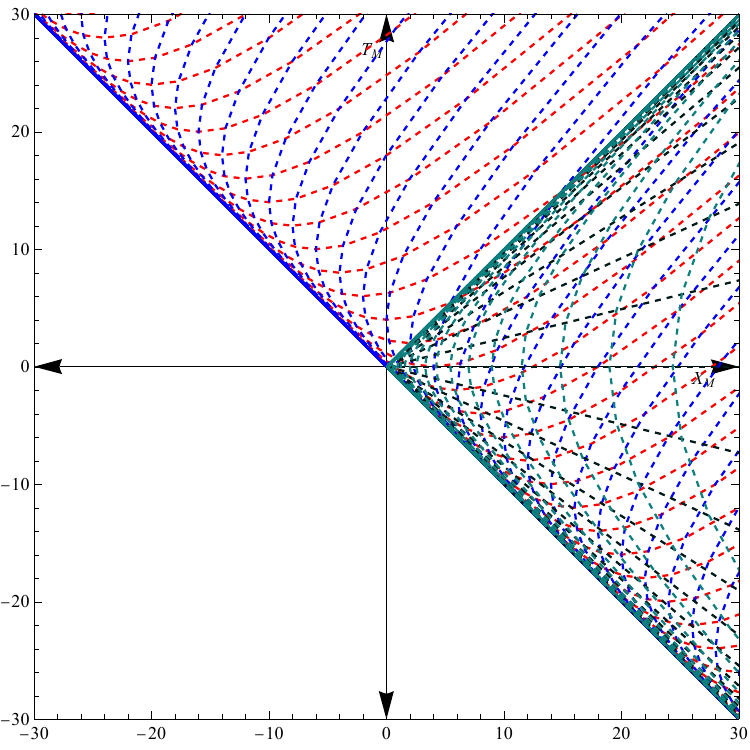}
    \caption{Half-sided left movers map: the spacetime has a thermal flux of left-moving particles}
    \label{Fig_2}
\end{figure}
We choose the coordinate transformation $V_M = \frac{1}{a} e^{a v_1}$ and $U_M = u_1$, as illustrated in Fig. ~\ref{Fig_2}. In terms of the new coordinates $(u_1, v_1)$, the metric can be expressed in the light-cone form
\begin{equation}
    ds^2 = -e^{a\;v_1} \;du_1\;dv_1 .  \label{Eq:5.1.0.15}
\end{equation}
 Equation~(\ref{Eq:5.1.0.15}) can be expressed in terms of the $(t_1, x_1)$ coordinates.
 \begin{equation}
     ds^2 =  - e^{a(t_1 + x_1)}\;(dt_1^2 - dx_1^2). \label{Eq:5.1.0.16} 
 \end{equation}
 The metric given in Eq.~(\ref{Eq:5.1.0.16}) and the spacetime diagram are illustrated in Fig. ~\ref{Fig_2}. The corresponding Bogoliubov coefficients are given by,
 \begin{equation}
   \alpha^*_{10}(\omega,\omega_0) = \frac{1}{2\;\pi\;a}\sqrt{\frac{\omega}{\omega_0}}\Gamma\bigg[\frac{i\omega}{a}\bigg]\,e^{\frac{\pi\,\omega}{2a}}\,\bigg(\frac{\omega_0}{a}\bigg)^{-\frac{i\,\omega}{a}}.  \label{Eq:5.1.0.17}   
 \end{equation}
 Similarly, we obtain
 \begin{equation}
    \beta^*_{10}(\omega,\omega_0) = -\frac{1}{2\;\pi\;a}\sqrt{\frac{\omega}{\omega_0}}\Gamma\bigg[\frac{i\omega}{a}\bigg]\,e^{-\frac{\pi\,\omega}{2a}}\,\bigg(\frac{\omega_0}{a}\bigg)^{-\frac{i\,\omega}{a}}. \label{Eq:5.1.0.18}
\end{equation}
The expectation value of the energy-momentum tensor is obtained after simplification as,
\begin{equation}
    \langle 0_M|:T_{v_1v_1}:| 0_M\rangle  = \frac{\hbar a^2}{48\pi},  \label{Eq:5.1.0.19}
\end{equation}
and,
\begin{equation}
    \langle 0_M|:T_{u_1u_1}:| 0_M\rangle  = 0.  \label{Eq:5.1.0.20}
\end{equation}

From Eqs.~(\ref{Eq:5.1.0.19}) and~(\ref{Eq:5.1.0.20}), we observe that the left-moving modes are excited, while the right-moving modes remain absent.
\subsubsection{Right Rindler spacetime}

\begin{figure}[h!]
    \centering
    \includegraphics[width=0.8\linewidth]{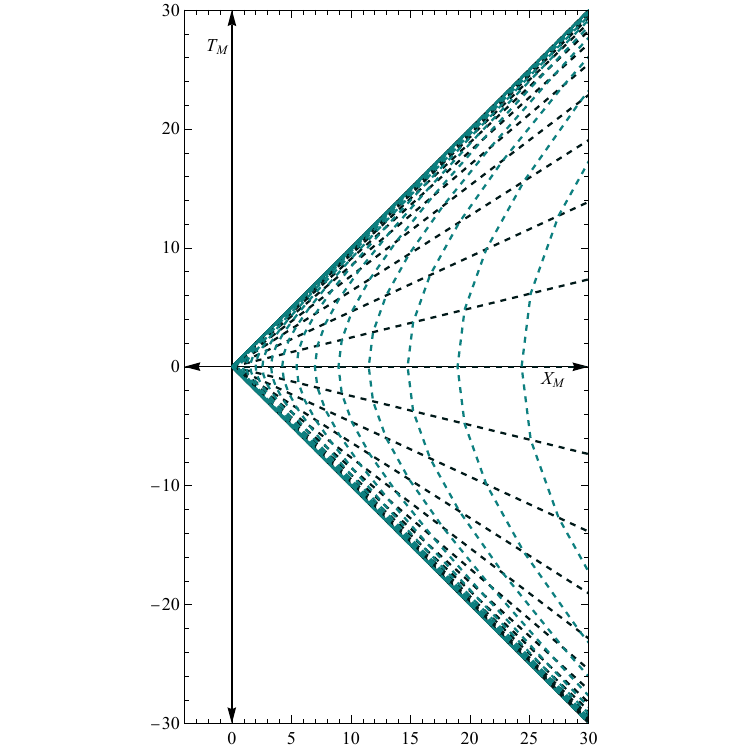}
    \caption{Right Rindler spacetime: This spacetime has a thermal density of left-moving and right-moving particles}
    \label{Fig_3}
\end{figure}
We choose the coordinate transformation,
\begin{equation}
    U_M = -\frac{1}{a} e^{-a u_1}, \qquad
    V_M = \frac{1}{a} e^{a v_1}. \label{Eq:5.1.0.21}
\end{equation}

Hence, the metric becomes,
\begin{equation}
    ds^2 = - e^{-a u_1} e^{a v_1} \, du_1 dv_1,\label{Eq:5.1.0.22}
\end{equation}
which in $(t_1,x_1)$ coordinates takes the form,
\begin{equation}
    ds^2 = - e^{2a x_1} (dt_1^2 - dx_1^2). \label{Eq:5.1.0.23}
\end{equation}
 The metric given in Eq.~(\ref{Eq:5.1.0.23}) is illustrated in Fig.~\ref{Fig_3}. The corresponding Schwarzian derivatives are given by,
\begin{equation}
    S(U_M,u_1) = S(V_M,v_1) = -\frac{a^2}{2}. \label{Eq:5.1.0.24}
\end{equation}

Thus, the expectation values of the stress-energy tensor are,
\begin{equation}
    \langle T_{u_1 u_1} \rangle = 
    \langle T_{v_1 v_1} \rangle 
    = \frac{\hbar a^2}{48\pi}. \label{Eq:5.1.0.25}
\end{equation}

This shows that both the right-moving and left-moving modes are thermally excited. The corresponding particle spectrum is Planckian with a temperature, 
\begin{equation}
    T = \frac{a}{2\pi}. \label{Eq:5.1.0.26}
\end{equation}

Hence, the Minkowski vacuum appears as a thermal state to a uniformly accelerated observer, which is the statement of the Unruh effect~\cite {Unruh:1976db, Unruh:1983ms, Birrell:1982ix, Frodden:2018mdm, Kolekar:2013hra}.
\subsubsection{Left Rindler spacetime}
\begin{figure}[h!]
    \centering
    \includegraphics[width=0.8\linewidth]{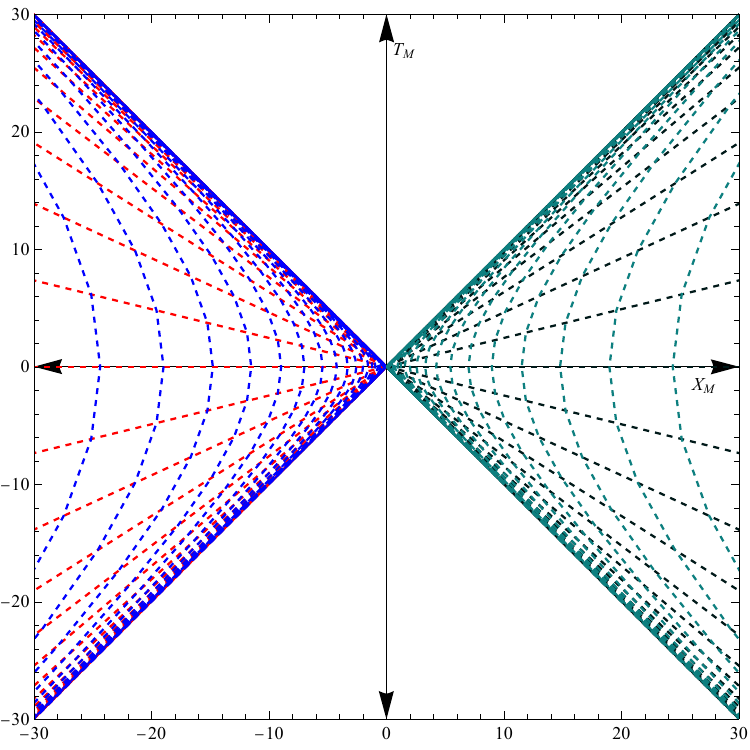}
    \caption{Left Rindler  (purple) along with Right Rindler wedge (green). The left wedge also has a thermal density of particles: both left-moving and right-moving.}
    \label{Fig_4}
\end{figure}
We consider the conformal transformation,
\begin{equation}
    U_M = \frac{1}{a} e^{a u_1}, \qquad 
    V_M = -\frac{1}{a} e^{-a v_1}.\label{Eq:5.1.0.27}
\end{equation}

Under this transformation, the metric becomes,
\begin{equation}
    ds^2 = - e^{a u_1} e^{-a v_1}  du_1 dv_1, \label{Eq:5.1.0.28}
\end{equation}
which in $(t_1,x_1)$ coordinates takes the form,
\begin{equation}
    ds^2 = - e^{-2a x_1} (dt_1^2 - dx_1^2). \label{Eq:5.1.0.29}
\end{equation}
The metric given in Eq.~(\ref{Eq:5.1.0.29}) is illustrated in Fig.~\ref{Fig_4}, which corresponds to the left Rindler wedge.

The Schwarzian derivatives for both null coordinates are identical,
\begin{equation}
    S(U_M,u_1) = S(V_M,v_1) = -\frac{a^2}{2}, \label{Eq:5.1.0.30}
\end{equation}
leading to
\begin{equation}
    \langle T_{u_1 u_1} \rangle = 
    \langle T_{v_1 v_1} \rangle 
    = \frac{\hbar a^2}{48\pi}. \label{Eq:5.1.0.31}
\end{equation}

Thus, both right-moving and left-moving modes are thermally excited with a Planckian spectrum at temperature,
\begin{equation}
    T = \frac{a}{2\pi}.\label{Eq:5.1.0.32}
\end{equation}

This spacetime describes the left Rindler wedge, which is causally disconnected from the right Rindler wedge. An observer confined to this region perceives the Minkowski vacuum as a thermal state, which is analogous to the standard Unruh effect in the right wedge. A reflection symmetry relates the two wedges and together forms a maximally extended description of uniformly accelerated observers.

\subsubsection{Future Milne universe}
We choose the maps $V_m=e^{av_1}/a$ and $U_M=e^{au_1}/a$.
\begin{figure}[h!]
    \centering
    \includegraphics[width=0.8\linewidth]{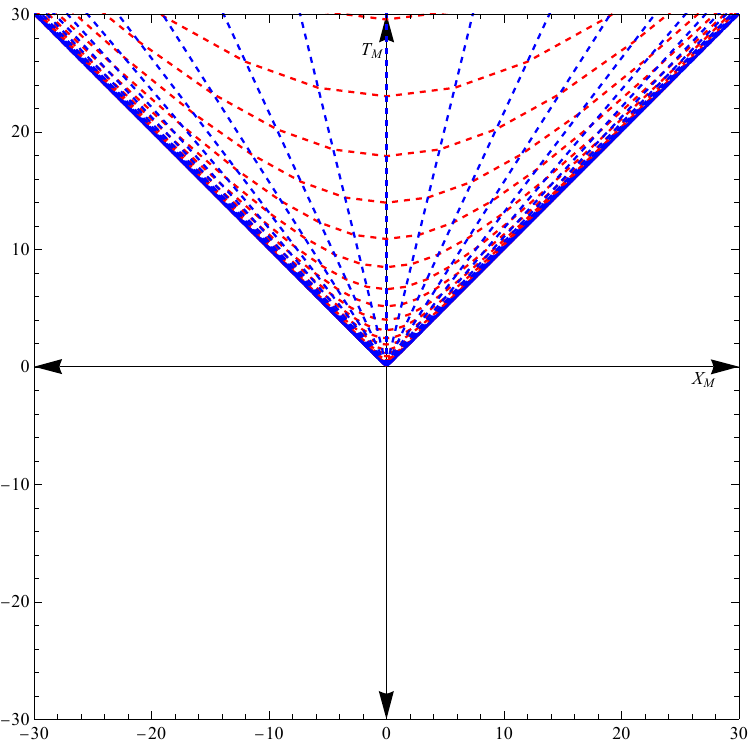}
    \caption{Future Milne universe}
    \label{Fig_5}
\end{figure}

Given a map into new coordinates(Future part of the Milne universe) in $(u_1,v_1)$, the Metric in these coordinates can be written in light-cone coordinates,
\begin{equation}
    ds^2 = -e^{a(u_1 + v_1)} du_1 dv_1. \label{Eq:5.1.0.33}
\end{equation}
Eq.~(\ref{Eq:5.1.0.33}) is written in terms of $(t_1,x_1)$-coordinates,
 \begin{equation}
     ds^2 = e^{2at_1}\big(-dt_1^2 + dx_1^2). \label{Eq:5.1.0.34}
 \end{equation}
 The metric given in Eq.~(\ref{Eq:5.1.0.34}), and the spacetime is the shaded portion in Fig. ~\ref{Fig_5}. In Milne coordinates, it becomes conformally flat with a scale factor $a(t)=e^{at}$. The corresponding conformally flat metric can be written as,
 \begin{equation}
     ds^2 = a^2(t)\big(-dt_1^2 + dx_1^2). \label{Eq:5.1.0.35}
 \end{equation}
 The Right-mover mode in the Milne coordinate is,
 \begin{equation}
    \hat{\phi}_{Mil.}(u_1) = \int_0^\infty \frac{d\Omega}{\sqrt{4\pi\Omega}}\bigg(\overrightarrow{\hat{c_1}}(\Omega) e^{-i\;\Omega\; u_1}  +\overrightarrow{\hat{c_1}}^\dagger(\Omega)e^{i\;\Omega\; u_1} \bigg) ,  \label{Eq:5.1.0.36}
\end{equation}
 and the left-mover mode is: 
 \begin{equation}
    \hat{\phi}_{Mil.}(v_1) = \int_0^\infty \frac{d\Omega}{\sqrt{4\pi\Omega}}\bigg(\overleftarrow{\hat{d_1}}(\Omega) e^{-i\;\Omega\; v_1}  +\overleftarrow{\hat{d_1}^\dagger}(\Omega)e^{i\;\Omega\; v_1} \bigg).  \label{Eq:5.1.0.37}
\end{equation}
 By simplifying using the Bogoliubov techniques in the Minkowski and Milne coordinates, we obtain the Bogoliubov coefficients as
 \begin{equation}
    \alpha^*(\Omega,\omega_0) = \frac{1}{2\;\pi\;a}\sqrt{\frac{\Omega}{\omega_0}}\bigg(\frac{a}{\omega_0}\bigg)^{\frac{i\;\Omega}{a}}\;e^{\frac{\pi \;\Omega}{2a}}\;\Gamma\bigg[\frac{i\Omega}{a}\bigg],  \label{Eq:5.1.0.38}
\end{equation}
and,
\begin{equation}
   \beta^*(\Omega,\omega_0) = -\frac{1}{2\;\pi\;a}\sqrt{\frac{\Omega}{\omega_0}}\bigg(\frac{a}{\omega_0}\bigg)^{\frac{-i\;\Omega}{a}}\;e^{-\frac{\pi \;\Omega}{2a}}\;\Gamma\bigg[-\frac{i\Omega}{a}\bigg].  \label{Eq:5.1.0.39} 
\end{equation}
Eqs.~(\ref{Eq:5.1.0.38}) and ~(\ref{Eq:5.1.0.39}) are the Bogoliubov coefficients for right-moving modes, and by symmetry, the left-moving modes yield the Bogoliubov transformation. Because the Metric is symmetric in $u_1$ and $v_1$.
From Eqs.~(\ref{Eq:5.1.0.38}) and ~(\ref{Eq:5.1.0.39}), we can write,
\begin{equation}
    \frac{\mid \beta\mid^2}{\mid \alpha\mid^2} = e^{-\frac{2\;\pi\;\Omega}{a}}. \label{Eq:5.1.0.40} 
\end{equation}
From Eq.~(\ref{Eq:5.1.0.40}), we obtained the Unruh temperature $T =\frac{a}{2\;\pi}$.
As calculated in the previous section, the component of the expectation value of stress-energy is calculated as,
\begin{equation}
    \langle 0_M|:T_{u_1u_1}:| 0_M\rangle  = \langle 0_M|:T_{v_1v_1}:| 0_M\rangle= \frac{\hbar a^2}{48\pi}.  \label{Eq:5.1.0.41} 
\end{equation}
From Eq.~(\ref{Eq:5.1.0.41}), we infer that both the left and right movers are excited.
\subsubsection{Past Milne universe}
We choose the map $V_m=-e^{-av_1}/a$ and $U_M=-e^{-au_1}/a$.
\begin{figure}[h!]
    \centering
    \includegraphics[width=0.8\linewidth]{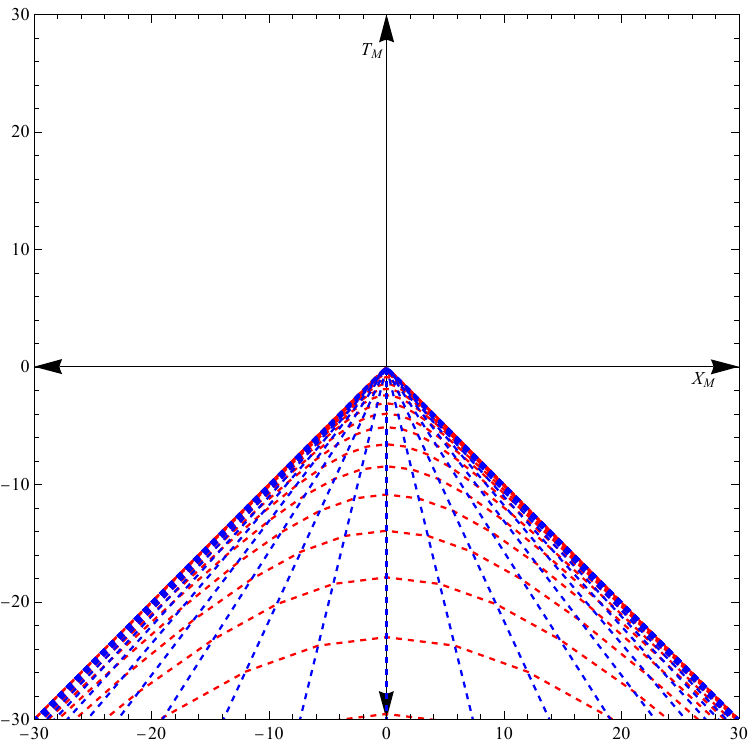}
    \caption{Past Milne universe}
    \label{Fig_6}
\end{figure}

Given a map into new coordinates(Past part of the Milne universe) in $(u_1,v_1)$, the Metric in these coordinates can be written in light-cone coordinates,
\begin{equation}
    ds^2 = -e^{-a(u_1 + v_1)} du_1 dv_1, \label{Eq:5.1.0.42}
\end{equation}
 Eq.~(\ref{Eq:5.1.0.42}) is written in terms of $(t_1,x_1)$-coordinates,
 \begin{equation}
     ds^2 = e^{-2at_1}\big(-dt_1^2 + dx_1^2). \label{Eq:5.1.0.43}
 \end{equation}
 The metric given in Eq.~(\ref{Eq:5.1.0.43}) is illustrated in Fig.~\ref{Fig_6}. As calculated, the Bogoliubov coefficient can be written as in the previous section,
 
 \begin{equation}
    \alpha^*(\Omega,\omega_0) = \frac{1}{2\;\pi\;a}\sqrt{\frac{\Omega}{\omega_0}}\bigg(\frac{\omega_0}{a}\bigg)^{\frac{i\;\Omega}{a}}\;e^{\frac{\pi \;\Omega}{2a}}\;\Gamma\bigg[-\frac{i\Omega}{a}\bigg].
    \label{Eq:5.1.0.44}
\end{equation}
Similarly,
\begin{equation}
   \beta^*(\Omega,\omega_0) = -\frac{1}{2\;\pi\;a}\sqrt{\frac{\Omega}{\omega_0}}\bigg(\frac{\omega_0}{a}\bigg)^{\frac{i\;\Omega}{a}}\;e^{-\frac{\pi \;\Omega}{2a}}\;\Gamma\bigg[-\frac{i\Omega}{a}\bigg], \label{Eq:5.1.0.45} 
\end{equation}
Eqs.~(\ref{Eq:5.1.0.44}) and ~(\ref{Eq:5.1.0.45}) are the Bogoliubov coefficients for right-moving modes, and by symmetry, the left-moving modes, we get the Bogoliubov transformation. Because the Metric is symmetric in $u_1$ and $v_1$.
From Eqs.~(\ref{Eq:5.1.0.44}) and ~(\ref{Eq:5.1.0.45}), we can write,
\begin{equation}
    \frac{\mid \beta\mid^2}{\mid \alpha\mid^2} = e^{-\frac{2\;\pi\;\Omega}{a}}. \label{Eq:5.1.0.46} 
\end{equation}
From Eq.~(\ref{Eq:5.1.0.46}), we obtain the Unruh temperature $T =\frac{a}{2\;\pi}$.
As calculated in the previous section, the component of the expectation value of stress-energy is calculated as,
\begin{equation}
    \langle 0_M|:T_{u_1u_1}:| 0_M\rangle  = \langle 0_M|:T_{v_1v_1}:| 0_M\rangle= \frac{\hbar a^2}{48\pi}.  \label{Eq:5.1.0.47} 
\end{equation}
From Eq.~(\ref{Eq:5.1.0.47}), we infer that both the left and right movers are excited.

\section{Maps starting from Rindler spacetime}
\label{sec-6_1}
In this section, we choose another starting point, viz., Rindler spacetime. We could have also chosen a Milne spacetime, but the Rindler spacetime is more interesting because it is the near-horizon geometry of black hole horizons. We can ask the same questions we posed for Minkowski spacetime and similarly obtain new spacetimes that are subsets of the Rindler spacetime with many interesting properties. One such spacetime has already been found in~\cite{Kolekar:2013hra}, where they have considered a Rindler-like transformation to the Rindler coordinates. They also showed that when one starts with Rindler spacetime in vacuum, the Rindler-Rindler spacetime has a thermal distribution of particles. In this section, we discover all possible subsets of the Rindler spacetime that have a thermal distribution of particles, of which the Rindler-Rindler spacetime is one among them. 

We choose $Y$ to be $u_1$ ( or $v_1$) and $X$ to be ($u_2$ (or $v_2$) in equation \ref{Eq:5.1.0.2}. We arrive at the following conformal maps
\subsection{Half-sided Rindler-Rindler spacetime: Right movers in excited state }
We choose the map $u_1=-e^{-au_2}/a$ and $v_1=v_2$.
\begin{figure}[h!]
    \centering
    \includegraphics[width=0.8\linewidth]{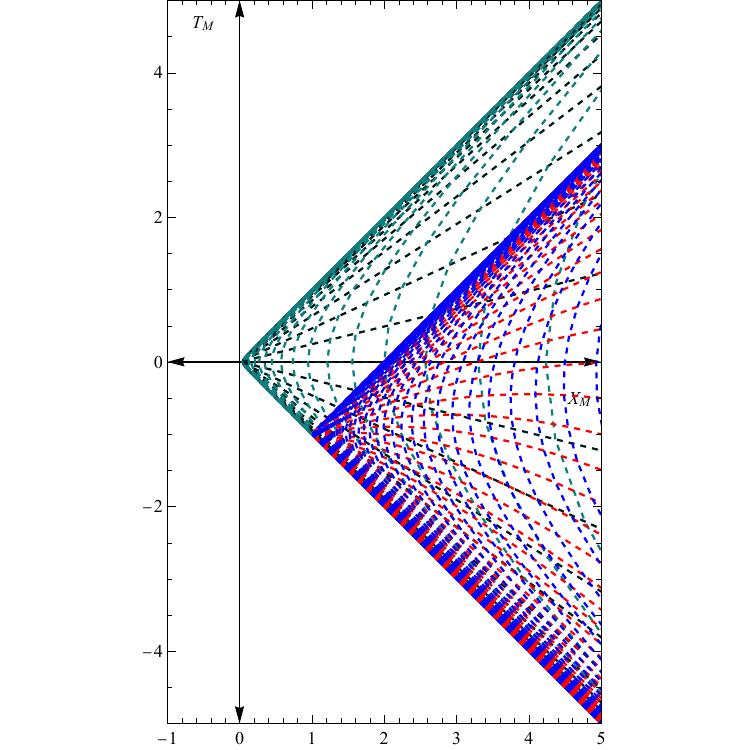}
    \caption{Half-sided Rindler-Rindler spacetime: Right movers in excited state}
    \label{Fig_7}
\end{figure}

The map in this section, with the $u$-sector, is exponentially changed, whereas the $v$-sector remains unchanged. The metric in the Rindler patch is calculated from  Eq.~(\ref{Eq:2.1.0.5}) is written as in light-cone coordinate,
\begin{equation}
    ds^2 = - e^{a(v_2 -u_2) +e^{-au_2}}\;du_2\;dv_2.  \label{Eq:5.1.0.48}
\end{equation}
The above metric is written in $(t_2,x_2)$ coordinate is,
\begin{equation}
     ds^2 = e^{2ax_2 + e^{-a(t_2-x_2)}} (-dt_2^2 +dx_2^2). \label{Eq:5.1.0.49}
\end{equation}
The metric given in Eq.~(\ref{Eq:5.1.0.49}) is illustrated in Fig.~\ref{Fig_7}. The green shaded area represents the Rindler spacetime, and the blue shaded region of a Rindler wedge is the half-sided Rindler-Rindler spacetime.  For the mapping under consideration, the left-moving sector satisfies ${v_1 = v_2}$. Upon simplification, the Bogoliubov coefficient is obtained as,
\begin{equation}
  \alpha^*(\nu,\omega) = \delta(\nu-\omega) \qquad  \beta^*(\nu,\omega) =0.\label{Eq:5.1.0.50}
\end{equation}
Similarly, the Bogoliubov coefficient for the right-moving mode is calculated as,
 \begin{equation}
    \alpha^*(\nu,\omega) = \frac{1}{2\;\pi\;a}\sqrt{\frac{\nu}{\omega}}\bigg(\frac{\omega}{a}\bigg)^{\frac{i\;\nu}{a}}\;e^{\frac{\pi \;\nu}{2a}}\;\Gamma\bigg[-\frac{i\nu}{a}\bigg] .  \label{Eq:5.1.0.51}
\end{equation}
Similarly,
\begin{equation}
    \beta^*(\nu,\omega) = -\frac{1}{2\;\pi\;a}\sqrt{\frac{\nu}{\omega}}\bigg(\frac{\omega}{a}\bigg)^{\frac{i\;\nu}{a}}\;e^{-\frac{\pi \;\nu}{2a}}\;\Gamma\bigg[-\frac{i\nu}{a}\bigg] .  \label{Eq:5.1.0.52}
\end{equation}
The thermal flux can be calculated as,
\begin{equation}
    \langle 0_{R_1}|:T_{u_2 u_2}:| 0_{R_1}\rangle  = \frac{\hbar a^2}{48\pi},  \label{Eq:5.1.0.53}
\end{equation}
and 
\begin{equation}
    \langle 0_{R_1}|:T_{v_2 v_2}:| 0_{R_1}\rangle  = 0.  \label{Eq:5.1.0.54}
\end{equation}
From the Eqs.~(\ref{Eq:5.1.0.53}) and ~(\ref{Eq:5.1.0.54}), the right-moving flux is present while the left movers are absent. This is again an instance of selective particle excitation. 
\subsubsection{Half-sided Rindler-Rindler spacetime: left mover flux}
We choose the map $v_1=e^{av_2}/a$ and $u_1=u_2$.
\begin{figure}[h!]
    \centering
    \includegraphics[width=0.8\linewidth]{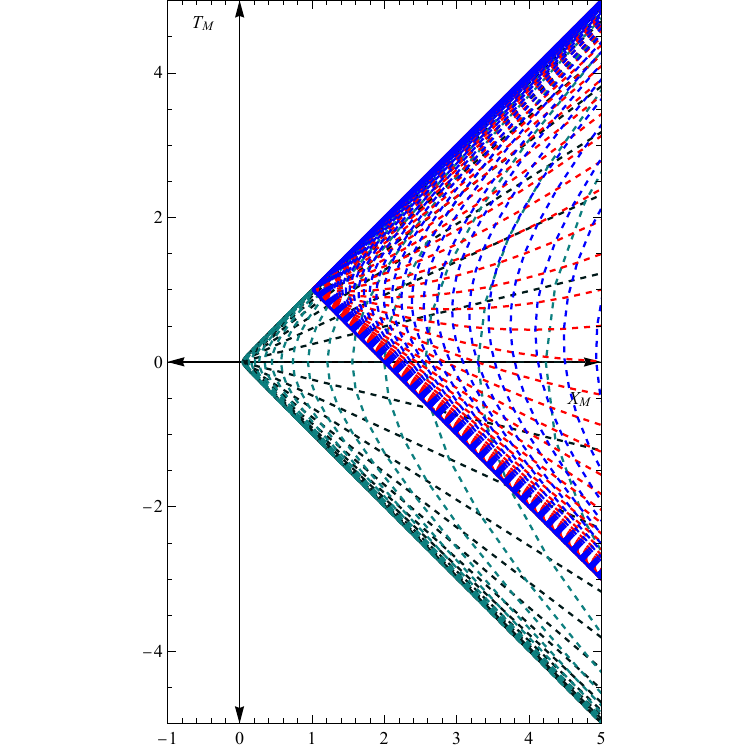}
    \caption{Half-sided Rindler-Rindler spacetime: left movers in excited state}
    \label{Fig_8}
\end{figure}

The map in this section, with the $v$-sector, is exponentially changed, while the $u$-sector remains unchanged. The metric in the Rindler patch is calculated from  Eq.~(\ref{Eq:2.1.0.5}) is written as in light-cone coordinate,
\begin{equation}
    ds^2 = - e^{e^{av_2}-a(u_2 - v_2)}\;du_2\;dv_2.  \label{Eq:5.1.0.55}
\end{equation}
The above metric is written in $(t_2,x_2)
$ coordinate is,
\begin{equation}
     ds^2 = e^{e^{a(t_2 + x_2)}+2ax_2} (-dt_2^2 +dx_2^2). \label{Eq:5.1.0.56}
\end{equation}

The metric given in Eq.~(\ref{Eq:5.1.0.56}) is shown in Fig. ~\ref{Fig_8}.The blue shaded region corresponds to the half-sided Rindler spacetime. For the mapping under consideration, the right-moving sector satisfies ${u_1 = u_2}$. Upon simplification, the Bogoliubov coefficient is obtained as,
\begin{equation}
  \alpha^*(\nu,\omega) = \delta(\nu-\omega), \qquad  \beta^*(\nu,\omega) =0.\label{Eq:5.1.0.57}
\end{equation}
Similarly, the Bogoliubov coefficient for the left-moving mode is calculated as,
 \begin{equation}
    \alpha^*(\nu,\omega) = \frac{1}{2\;\pi\;a}\sqrt{\frac{\nu}{\omega}}\bigg(\frac{\omega}{a}\bigg)^{\frac{i\;\nu}{a}}\;e^{\frac{\pi \;\nu}{2a}}\;\Gamma\bigg[-\frac{i\nu}{a}\bigg] .  \label{Eq:5.1.0.58}
\end{equation}
Similarly,
\begin{equation}
    \beta^*(\nu,\omega) = -\frac{1}{2\;\pi\;a}\sqrt{\frac{\nu}{\omega}}\bigg(\frac{\omega}{a}\bigg)^{\frac{i\;\nu}{a}}\;e^{-\frac{\pi \;\nu}{2a}}\;\Gamma\bigg[-\frac{i\nu}{a}\bigg] .  \label{Eq:5.1.0.59}
\end{equation}
The thermal flux can be calculated as,
\begin{equation}
    \langle 0_{R_1}|:T_{v_2 v_2}:| 0_{R_1}\rangle  = \frac{\hbar a^2}{48\pi},  \label{Eq:5.1.0.60}
\end{equation}
and;
\begin{equation}
    \langle 0_{R_1}|:T_{u_2 u_2}:| 0_{R_1}\rangle  = 0.  \label{Eq:5.1.0.61}
\end{equation}
From the Eqs.~(\ref{Eq:5.1.0.60}) and ~(\ref{Eq:5.1.0.61}), the left-movers are in a thermal distribution, whereas there are no right-movers. 
\subsubsection{Right Rindler-Rindler spacetime}
 
\begin{figure}[h!]
    \centering
    \includegraphics[width=0.8\linewidth]{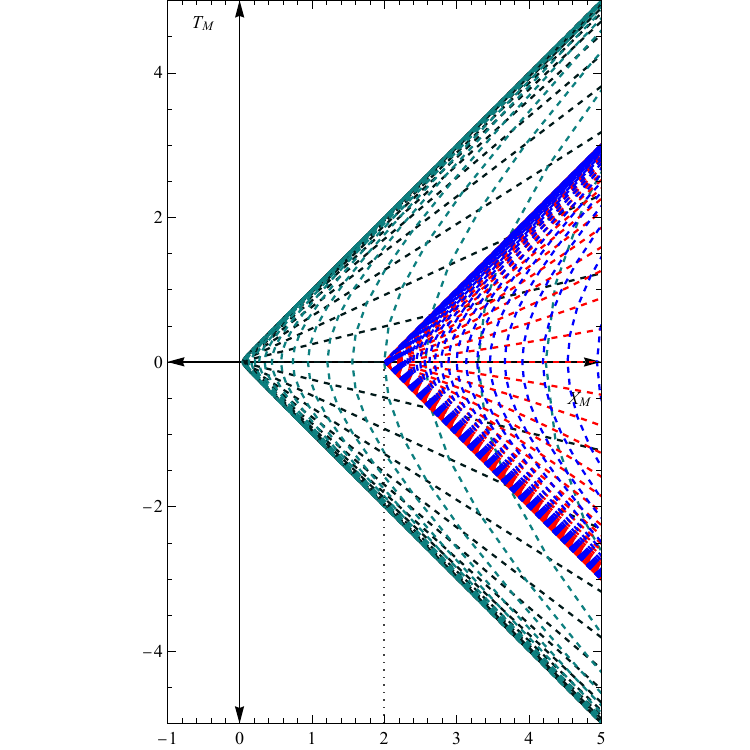}
    \caption{Rindler-Rindler spacetime: Thermal density of left movers and right movers}
    \label{Fig_9}
\end{figure}
By considering the conformal transformation,
\begin{equation}
    u_1 = -\frac{1}{a} e^{-a u_2}, \qquad 
    v_1 = \frac{1}{a} e^{a v_2}. \label{Eq:5.1.0.62}
\end{equation}

Under this transformation, the Rindler metric,
\begin{equation}
    ds^2 = - e^{a(v_1 - u_1)} \, du_1 dv_1 .  \label{Eq:5.1.0.63}
\end{equation}
Takes the form,
\begin{equation}
    ds^2 = - e^{a\left(\frac{1}{a}e^{a v_2} + \frac{1}{a}e^{-a u_2}\right)} 
    e^{-a u_2} e^{a v_2} \, du_2 dv_2.  \label{Eq:5.1.0.64}
\end{equation}

In $(t_2,x_2)$ coordinates, this corresponds to a highly nontrivial conformal factor,
\begin{equation}
    ds^2 = - \exp\!\left[2a x_2 + e^{a(t_2 + x_2)} + e^{-a(t_2 - x_2)}\right]
    (dt_2^2 - dx_2^2).  \label{Eq:5.1.0.65}
\end{equation}
The metric given in Eq.~(\ref{Eq:5.1.0.65}) is shown in Fig. ~\ref{Fig_9}. The blue shaded region that has the shape of a diamond corresponds to the left Rindler-Rindler spacetime. This spacetime is found in \cite{Kolekar:2013hra}.  In this case, both the null sectors undergo nontrivial exponential transformations. Consequently, the Bogoliubov coefficients for both the right-moving and left-moving modes are non-vanishing, indicating particle production in both sectors. The corresponding expectation values of the stress-energy tensor are given by,
\begin{equation}
    \langle 0_{R_1} | :T_{u_2 u_2}: | 0_{R_1} \rangle 
    = 
    \langle 0_{R_1} | :T_{v_2 v_2}: | 0_{R_1} \rangle 
    = \frac{\hbar a^2}{48\pi}, \label{Eq:5.1.0.66}
\end{equation}
implying that both right-moving and left-moving modes are thermally excited.

 The resulting particle distribution is again Planckian with temperature
\begin{equation}
    T = \frac{a}{2\pi}. \label{Eq:5.1.0.67}
\end{equation}

A detailed derivation of the Bogoliubov coefficients and their thermal interpretation follows standard methods and can be found in  \cite{Kolekar:2013hra}. 
\subsubsection{Left Rindler-Rindler Diamond spacetime}

\begin{figure}[h!]
    \centering
    \includegraphics[width=0.8\linewidth]{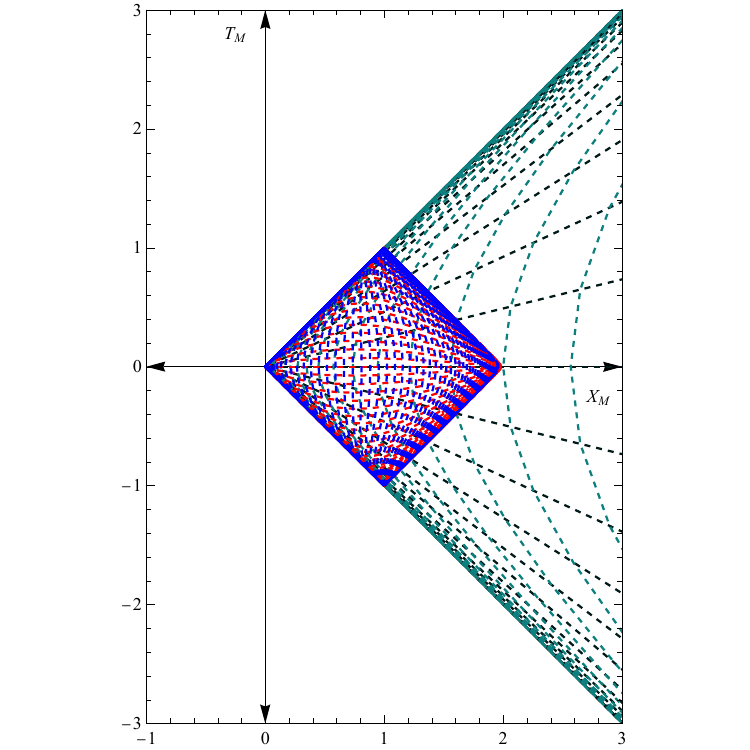}
    \caption{Left Rindler-Rindler Diamond : Has both left movers and right movers with thermal spectrum}
    \label{Fig_10}
\end{figure}
We consider the conformal transformation,
\begin{equation}
    u_1 = \frac{1}{a} e^{a u_2}, \qquad 
    v_1 = -\frac{1}{a} e^{-a v_2}.\label{Eq:5.1.0.68}
\end{equation}

Under this transformation, the Rindler metric,
\begin{equation}
    ds^2 = - e^{a(v_1 - u_1)} \, du_1 dv_1. \label{Eq:5.1.0.69}
\end{equation}
Which is,
\begin{equation}
    ds^2 = - e^{-\left(e^{-a v_2} +e^{a u_2}\right)} 
    e^{a u_2} e^{-a v_2} \, du_2 dv_2, \label{Eq:5.1.0.70}
\end{equation}

In $(t_2,x_2)$ coordinates, the metric takes the form,
\begin{equation}
    ds^2 = - \exp\!\left[-2a x_2 - e^{a(t_2 - x_2)} - e^{-a(t_2 + x_2)}\right]
    (dt_2^2 - dx_2^2).\label{Eq:5.1.0.71}
\end{equation}
The metric given in Eq.~(\ref{Eq:5.1.0.71}) is illustrated in Fig.~\ref{Fig_10}. In this case, both null sectors undergo nontrivial exponential transformations. Consequently, the Bogoliubov coefficients for the right-moving and left-moving modes are non-vanishing, indicating particle production in both sectors. The corresponding expectation values of the stress-energy tensor are given by,
\begin{equation}
    \langle 0_{R_1} | :T_{u_2 u_2}: | 0_{R_1} \rangle 
    = 
    \langle 0_{R_1} | :T_{v_2 v_2}: | 0_{R_1} \rangle 
    = \frac{\hbar a^2}{48\pi}, \label{Eq:5.1.0.72}
\end{equation}

The corresponding Bogoliubov coefficients yield a Planckian particle distribution,
\begin{equation}
    \langle N_\nu \rangle = \frac{1}{e^{2\pi \nu/a} - 1},\label{Eq:5.1.0.73}
\end{equation}
Indicating a thermal spectrum at a temperature,
\begin{equation}
    T = \frac{a}{2\pi}, \label{Eq:5.1.0.74}
\end{equation}

This spacetime represents a left Rindler-Rindler diamond, which is causally disconnected from its right counterpart. An observer confined to this region perceives the Rindler vacuum as a thermal state, analogous to the Unruh effect, but now arising from a successive conformal transformation within the Rindler wedge.
\subsubsection{Future Rindler-Milne universe}
 We choose the map $v_1=e^{av_2}/a$ and $u_1=e^{au_2}/a$.
\begin{figure}[h!]
    \centering
    \includegraphics[width=0.8\linewidth]{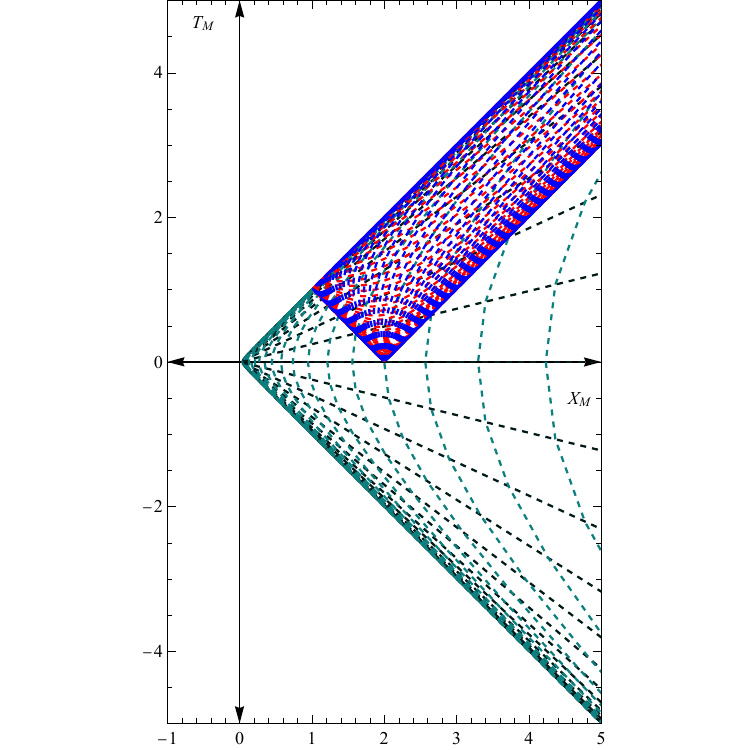}
    \caption{Future Rindler-Milne universe}
    \label{Fig_11}
\end{figure}

The Metric for the given map in this section, in light-cone coordinates,
\begin{equation}
    ds^2 = - e^{(e^{a v_2} -e^{au_2})}\; e^{a(u_2 + v_2)}du_2dv_2,\label{Eq:5.1.0.75}
\end{equation}
 The above metric is written in $(t_2,x_2)$ coordinate is,
\begin{equation}
     ds^2 = e^{(e^{a(t_2 + x_2)}-e^{a(t_2-x_2)})} e^{2at_2} (-dt_2^2 +dx_2^2), \label{Eq:5.1.0.76}
\end{equation}
The metric given in Eq.~(\ref{Eq:5.1.0.76}) is illustrated in Fig.~\ref{Fig_11}. Following the same procedure as in the previous section, the Bogoliubov coefficient is obtained as,
 \begin{equation}
    \alpha^*(\nu,\omega) = \frac{1}{2\;\pi\;a}\sqrt{\frac{\nu}{\omega}}\bigg(\frac{\omega}{a}\bigg)^{\frac{i\;\nu}{a}}\;e^{\frac{\pi \;\nu}{2a}}\;\Gamma\bigg[-\frac{i\nu}{a}\bigg] ,  \label{Eq:5.1.0.77}
\end{equation}
and,
\begin{equation}
    \beta^*(\nu,\omega) = -\frac{1}{2\;\pi\;a}\sqrt{\frac{\nu}{\omega}}\bigg(\frac{\omega}{a}\bigg)^{\frac{i\;\nu}{a}}\;e^{-\frac{\pi \;\nu}{2a}}\;\Gamma\bigg[-\frac{i\nu}{a}\bigg] .  \label{Eq:5.1.0.78}
\end{equation}
The Eqs.~(\ref{Eq:5.1.0.77}) and ~(\ref{Eq:5.1.0.78}) are the Bogoliubov coefficients for right-moving modes, and by symmetry, the left-moving modes, we get the same Bogoliubov transformation. This is because the metric is symmetric in $u_2$ and $v_2$.
As calculated in the previous section, the component of the expectation value of stress-energy is calculated as,
\begin{equation}
    \langle 0_{R_1}|:T_{u_2u_2}:| 0_{R_1}\rangle  = \langle 0_{R_1}|:T_{v_2v_2}:| 0_{R_1}\rangle= \frac{\hbar a^2}{48\pi}.  \label{Eq:5.1.0.79} 
\end{equation}
From Eq.~(\ref{Eq:5.1.0.79}), we infer that the left and right movers are excited.
\subsubsection{Past Rindler-Milne universe}
We choose the map $v_1=-e^{-av_2}/a$ and $u_1=-e^{-au_2}/a$.
\begin{figure}[h!]
    \centering
    \includegraphics[width=0.8\linewidth]{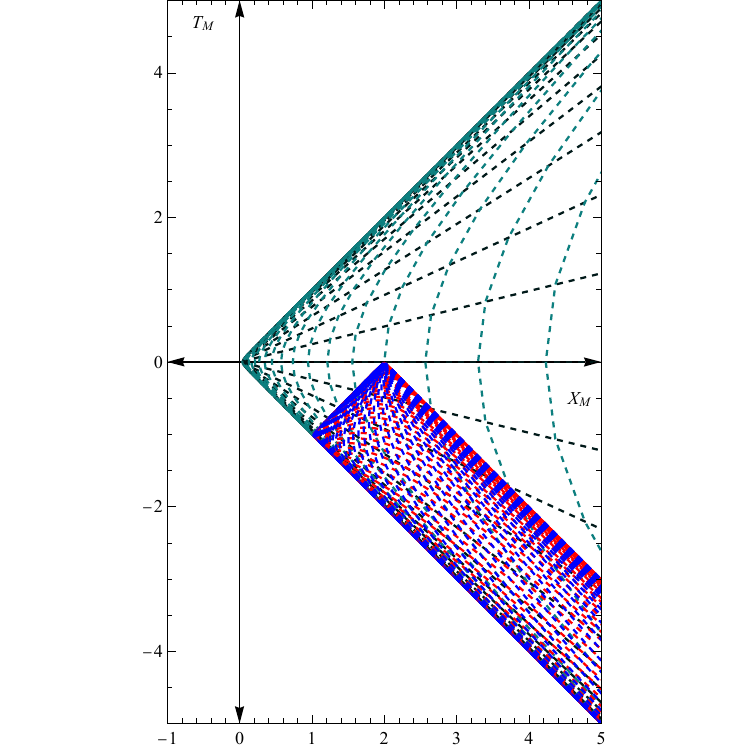}
    \caption{Past Rindler-Milne universe}
    \label{Fig_12}
\end{figure}

The Metric for the given map in this section, in light-cone coordinates,
\begin{equation}
    ds^2 = - e^{(e^{-a v_2} -e^{-au_2})}\; e^{-a(u_2 + v_2)}du_2dv_2,\label{Eq:5.1.0.80}
\end{equation}
 The above metric is written in $(t_2,x_2)$ coordinate is,
\begin{equation}
     ds^2 = e^{(e^{-a(t_2 - x_2)}-e^{-a(t_2+x_2)})} e^{-2at_2} (-dt_2^2 +dx_2^2), \label{Eq:5.1.0.81}
\end{equation}
The metric given in Eq.~(\ref{Eq:5.1.0.81}) is illustrated in Fig.~\ref{Fig_12}. Following the same procedure as in the previous section, the Bogoliubov coefficient is obtained as,
 \begin{equation}
    \alpha^*(\nu,\omega) = \frac{1}{2\;\pi\;a}\sqrt{\frac{\nu}{\omega}}\bigg(\frac{\omega}{a}\bigg)^{\frac{i\;\nu}{a}}\;e^{\frac{\pi \;\nu}{2a}}\;\Gamma\bigg[-\frac{i\nu}{a}\bigg] ,  \label{Eq:5.1.0.82}
\end{equation}
and,
\begin{equation}
    \beta^*(\nu,\omega) = -\frac{1}{2\;\pi\;a}\sqrt{\frac{\nu}{\omega}}\bigg(\frac{\omega}{a}\bigg)^{\frac{i\;\nu}{a}}\;e^{-\frac{\pi \;\nu}{2a}}\;\Gamma\bigg[-\frac{i\nu}{a}\bigg] .  \label{Eq:5.1.0.83}
\end{equation}
The Eqs.~(\ref{Eq:5.1.0.82}) and ~(\ref{Eq:5.1.0.83}) are the Bogoliubov coefficients for both sectors. Because the Metric is symmetric in $u_2$ and $v_2$.
As calculated in the previous section, the component of the expectation value of stress-energy is calculated as,
\begin{equation}
    \langle 0_{R_1}|:T_{u_2u_2}:| 0_{R_1}\rangle  = \langle 0_{R_1}|:T_{v_2v_2}:| 0_{R_1}\rangle= \frac{\hbar a^2}{48\pi}.  \label{Eq:5.1.0.84} 
\end{equation}
From Eq.~(\ref{Eq:5.1.0.84}), we infer that the left mover as well as the right mover are excited.
\section{Reconstruction of Vacuum Spacetimes From Excited States\label{sec-7_1}}

We consider the inverse problem of determining whether a spacetime with a non-vanishing energy flux can be mapped to a vacuum configuration by a suitable coordinate transformation.

Let us assume that we are in a given spacetime with null coordinates $(u,v)$ and constant fluxes $T_{uu}=\alpha^2$ and/or $T_{vv}=\alpha^2$. We want to find out the properties of a spacetime with the null 
coordinates $(U_p, V_p)$. The maps $u(U_p)$ and $v(V_p)$ will be such that the stress tensor in the $(U_p, V_P)$ is zero.
\begin{equation}
     \frac{1}{ (\frac{\partial u}{\partial U_p} )^2}\left[(\frac{\partial u}{\partial U_p} \left(\frac{\partial^3u}{\partial U_p^3}\right)-\frac{3}{2}\left(\frac{\partial^2 u}{\partial U_p^2}\right)^2\right]-\alpha^2(\frac{\partial u}{\partial U_p} )^2 =0, \label{Eq:6.1.0.1} 
\end{equation}
The general solution for the above is to see Appendix~\ref {Apn3},
\begin{equation}
u=\frac{1}{\alpha}ln\left| \frac{aU_p+b}{cU_p+d} \right|
, \label{Eq:6.1.0.2} 
\end{equation}
A similar expression for the $v$ sector,
\begin{equation}
v=\frac{1}{\alpha}ln\left| \frac{pV_p+q}{rV_p+s} \right|.
 \label{Eq:6.1.0.3} 
\end{equation}
The above transformation corresponds to the exponential map characteristic of the Rindler-type coordinates. We suppose that the Metric in the given spacetime is given by

\begin{equation}
    ds^2=-\Phi(u,v)dudv.  \label{Eq:6.1.0.4} 
\end{equation}
Using Eqs.~(\ref{Eq:6.1.0.2}) and (\ref{Eq:6.1.0.3}), we can obtain the metric of the purifying spacetime that is in vacuum. We note that we can purify sector by sector. This implies that we can purify the left-moving sector while preserving the flux of the right-moving sector, and vice versa. Thus, both sectors can be purified using this method. We note that the map gives the metric of the purifying spacetime that is valid in the given spacetime. The range can be extended to obtain the full spacetime.  
\subsection{Partial Purification of Diamond spacetime}

To make the partial purification procedure explicit, we demonstrate how a diamond spacetime can be mapped to the Rindler wedge.

We begin with a metric of the diamond spacetime in the form,
\begin{equation}
ds^2 = - e^{-a( v_2- u_2)}\left[e^{-(e^{a u_2 }+e^{-a v_2})}\right] \, du_2 dv_2.  \label{Eq:6.1.0.18}
\end{equation}

We perform a partial purification by transforming only the $u$-sector while keeping the $v$-sector untouched. We note that this transformation is not unique, and one may choose various constants $p,q,r,s$ in \ref{Eq:6.1.0.3} to generate families of purifying spacetimes. We present below the procedure 
\begin{equation}
U_p = \frac{e^{a u_2}}{a}, \qquad dU_p = e^{a u_2} du_2. \label{Eq:6.1.0.19}
\end{equation}

This implies,
\begin{equation}
du_2 = e^{-a u_2} dU_p.  \label{Eq:6.1.0.20}
\end{equation}

Substituting into the metric ~(\ref{Eq:6.1.0.18}), we obtain
\begin{equation}
ds^2 
= - e^{-a(v_2-U_p)}\, e^{-e^{-a v_2}}   dU_p dv_2. \label{Eq:6.1.0.21}
\end{equation}
Similarly, we can partially purify the other sector, leaving the left-moving sector the metric corresponding to the map,

Next, we redefine the $v$-sector coordinate as,
\begin{equation}
V_p = -\frac{e^{-a v_2}}{a}, \qquad dV_p = e^{-a v_2} dv_2.  \label{Eq:6.1.0.23}
\end{equation}

Substituting this into the metric ~(\ref{Eq:6.1.0.18}),
\begin{equation}
    ds^2  = - e^{a( u_2+V_p)} e^{-e^{a u_2}} du_2 dV_p.  \label{Eq:6.1.0.24}
\end{equation}
\begin{figure}[ht!]
    \centering
    \includegraphics[width=0.8\linewidth]{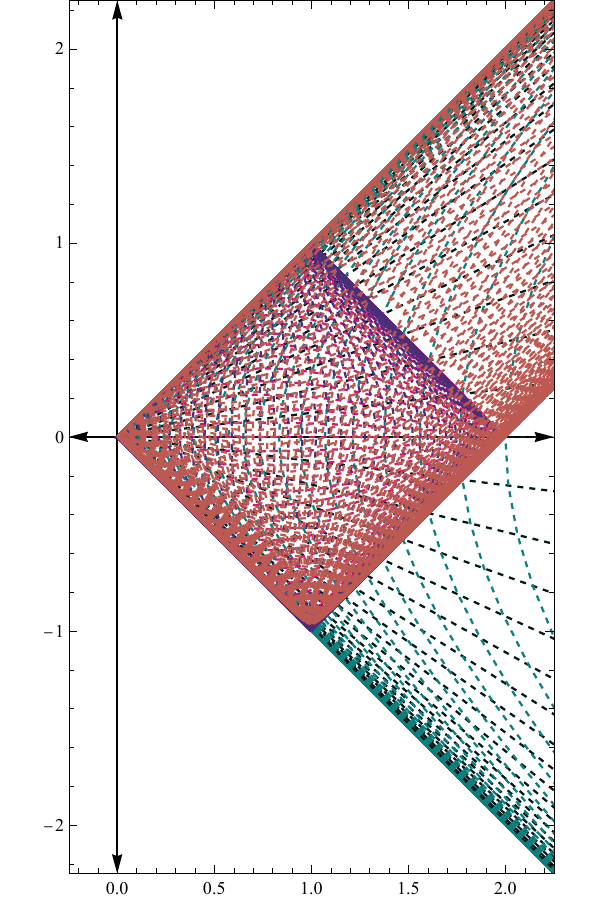}
    \caption{Partial purification of the Diamond spacetime: The original Diamond spacetime has a thermal density of particles in both sectors. The shaded strip is the partially purifying spacetime that has no flux of particles left moving particles}
     \label{Fig_14}
\end{figure}
Metrics ~(\ref{Eq:6.1.0.21}) and ~(\ref{Eq:6.1.0.24}) represent the geometries obtained after partial purification in one of the null sectors. The spacetime for the metric ~(\ref{Eq:6.1.0.24}) is illustrated in Fig.~\ref{Fig_14}. We note that the same spacetime can be obtained from the Rindler spacetime using the map. The remaining exponential factors encode the residual conformal structure inherited from the original diamond geometry.
\subsection{Full purification of Diamond spacetime}
We can carry out a complete purification of the diamond spacetime by carrying out the transformation by choosing $p,q,r,s$ for both the sectors. We illustrate one such choice, which is a canonical one in the context of Rindler spacetime. To make the Full purification procedure explicit, we choose,
\begin{equation}
V_p = -\frac{e^{-a v_2}}{a}, \qquad U_p = \frac{e^{a u_2}}{a}. \label{Eq:6.1.0.25}
\end{equation}
Substituting  Eq.~(\ref{Eq:6.1.0.25})into Eq.~(\ref{Eq:6.1.0.21}), we obtain,
\begin{equation}
    ds^2 = - e^{a(V_P-U_P)}\, dV_P dU_P.\label{Eq:6.1.0.26}
\end{equation}
We recognize this spacetime as the Rindler spacetime. Thus, we can reconstruct a fully purifying spacetime in this way. The purifying spacetime of a diamond spacetime with a thermal density of particles is a Rindler spacetime. 

\section{Conformal Maps Preserving Particle content: Sibling spacetimes\label{sec-8_1}}
In the case of Minkowski spacetime with a field in the vacuum state, a Rindler wedge detects the thermal density of particles. Now the bifurcation point of a Rindler wedge can be located anywhere in the spacetime. Two Rindler wedges that do not share the same bifurcation points but nevertheless share the same particle content owing to the same ``parent'' spacetime can be termed as sibling spacetimes. \\ 
We ask a related question. Given a spacetime with a density or flux of radiation, can we locate its sibling spacetimes? These are spacetimes that share the same parent spacetime and have the same particle content derived from the parent spacetime. We can locate the siblings by demanding that the flux remains constant upon transformation to the new sibling coordinates. 
Consider a spacetime with null coordinates $(u_1, v_1)$ and constant flux
\begin{equation}
T_{u_1 u_1} = \alpha^2, \qquad T_{v_1 v_1} = \alpha^2.
\end{equation}
We seek transformations $(u_1, v_1) \to (u_2, v_2)$ that leave this flux unaltered.

Using the anomalous transformation law of the stress tensor, this requirement leads to the Schwarzian equation
\begin{equation}
\begin{split}
\frac{1}{\left(\frac{\partial u_1}{\partial u_2}\right)^2}
\left[
\left(\frac{\partial u_1}{\partial u_2}\right)
\left(\frac{\partial^3u_1}{\partial u_2^3}\right)
-\frac{3}{2}
\left(\frac{\partial^2u_1}{\partial u_2^2}\right)^2
\right]\\
-\frac{\alpha^2}{2}
\left(\frac{\partial u_1}{\partial u_2}\right)^2
= -\frac{\alpha^2}{2}, 
\end{split}
\label{Eq:7.0.0.1}
\end{equation}
with an identical equation for the $v$-sector.

The general solution is obtained by the method given in Appendix~\ref {Apn3},
\begin{equation}
e^{\alpha u_1} = \frac{A e^{\alpha u_2} + B}{C e^{\alpha u_2} + D}, 
\qquad AD - BC \neq 0.\label{Eq:7.0.0.2}
\end{equation}
Which corresponds to a Möbius transformation acting on the exponential coordinate $e^{\alpha u}$. An ingoing sector produces the same equations with $u$ replaced by $v$. These transformations generate a class of coordinate systems with identical particle contents, starting from the same parent. We note that in this method, we need not solve for the parent spacetime and obtain the sibling spacetime directly by solving a differential equation. We illustrate the above method in the interesting case of the diamond spacetime.

\subsection{Diamond siblings}

\begin{figure}[ht!]
    \centering
    \includegraphics[width=0.8\linewidth]{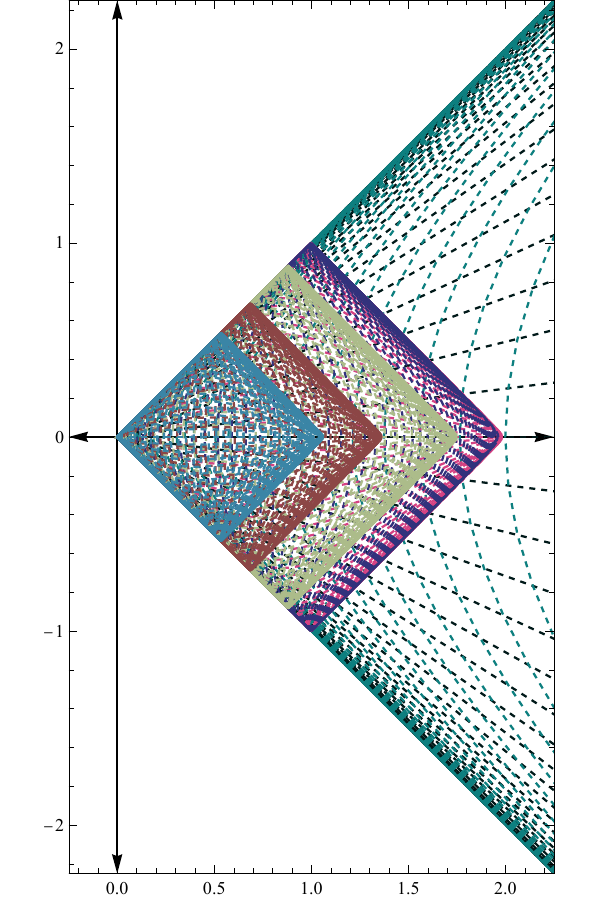}
    \caption{ Causal diamond siblings: Each of the diamond has both left movers and right movers in a thermal distribution}
     \label{Fig_13}
\end{figure}

We chose a diamond spacetime with a thermal density of particles. Based on the purification section, we note that this implies that the parent spacetime is a Rindler spacetime. However, the analysis in this section does not require us to determine the purifying spacetime. We now seek which other spacetimes overlap with this spacetime and are siblings of the diamond spacetime. We choose $u_3,v_3$ to be the null coordinates of the sibling spacetime that we are interested in and $u_2,v_2$ to be the coordinates of the Diamond spacetime with the metric given by \ref{Eq:6.1.0.18}. The coordinates $(u_1,v_1)$ are the null coordinates of the parent Rindler spacetime, as shown in green in figure \ref{Fig_13}.  Using  Eq.~(\ref{Eq:7.0.0.2}) we choose the following map (by setting the parameters $A,B,C,D$) in equation \ref{Eq:7.0.0.2}.
\begin{equation}
e^{au_3} = e^{au_2}-\Delta, \label{Diamondu}
\end{equation}
and 
\begin{equation}
e^{-av_3} = e^{-av_2}-\Delta, \label{Diamondv}
\end{equation}
with $\Delta$ a positive constant. One can choose independent sets of $\Delta$s in the above equations.  We choose the same $\Delta$ as they generate an aesthetically better-looking set of siblings in figure \ref{Fig_13}. We note that the metric of the first diamond and its sibling differ by an overall factor $e^{-2\Delta}$.
We change $\Delta$ and obtain various diamond spacetimes. A few of these are depicted in figure \ref{Fig_13}. All these diamonds have a thermal density of particles in both the left-moving and right-moving sectors when the parent spacetime is in vacuum. 

\section{Discussion and Conclusions}
In this article, we have generated and defined new spacetimes serving various roles in terms of particle content and their relationships with other spacetimes. We have generated the spacetimes for the three questions based on the Schwarzian expression and the corresponding differential equations they generate. Although we have a general solution based on this approach, the formal question regarding whether the obtained general solution is inclusive of all possible solutions for the given problem has yet to be established. We also note that not all the starting points discussed in the article are physically well-defined. For instance, a few maps start from a Rindler spacetime in vacuum, which is not a globally well-defined quantity, similar to a Boulware vacuum in a black hole spacetime. In this sense, the calculations done in this article are well defined in a local sense. However, we need the information on how the Rindler vacuum relates to the particle content of its subsets when one studies aspects when horizons evolve \cite{Kolekar:2013hra}, \cite{Lochan:2025mru}, \cite{jha2026inequivalentpathsthermalityminkowski}, \cite{Jha:2025tpg}, \cite{Gutti:2022xov}. \par
An interesting question based on this work is what happens when we replace the massless scalar field with a massive one. Based on the work, we have generated various spacetimes with a massless scalar field in the corresponding particle content. In the case of a massive field, the split into the left-moving and right-moving sectors is absent because the mass couples both sectors. What happens to thermality and particle content for massive fields is a question we leave for future considerations. \par
A useful question is how these spacetimes found in two dimensions generalize to four dimensions. What happens to the questions regarding particle content when we take the horizon structures for various spacetimes found in this study and add two more orthogonal directions for both massless and massive cases? Since the Schwarzian approach is valid only in two dimensions, how can one generalize the method of generating and purifying spacetimes in four dimensions? These questions, too, are taken up elsewhere. \par

\appendix
\section{Direct Solution via Schwarzian Equation\label{Apn1}}

We now solve the Schwarzian equation directly,
\begin{equation}
\{Y,X\} = -a^2, \label{Apn1.1}
\end{equation}

Writing this explicitly,
\begin{equation}
\frac{Y'''}{Y'} - \frac{3}{2} \left(\frac{Y''}{Y'}\right)^2 = -a^2,
\label{Apn1.2}
\end{equation}
and defining $p = Y'$, we obtain
\begin{equation}
\frac{p''}{p} - \frac{3}{2} \left(\frac{p'}{p}\right)^2 = -a^2, \label{Apn1.3}
\end{equation}

Introducing $\omega = \ln p$, this reduces to
\begin{equation}
\omega'' - \frac{1}{2} (\omega')^2 = -a^2, \label{Apn1.4}
\end{equation}

Setting $\omega' = v$, we obtain a Riccati equation,
\begin{equation}
v' - \frac{1}{2} v^2 = -a^2,  \label{Apn1.5}
\end{equation}

Solving this equation~\cite{riveraoliva2025solvingriccatiequation} and integrating back yields
\begin{equation}
Y(X) = \frac{A e^{aX} + B}{C e^{aX} + D}, \label{Apn1.6}
\end{equation}
with $AD - BC \neq 0$.
\section{Bogolyobov calculation\label{Apn2}}
We compute the Bogoliubov coefficients using the Klein–Gordon inner product evaluated on a null hypersurface, which we choose to be $V_M = \text{const}$ and  $U_M = \text{const}$. Although this surface is not a Cauchy surface in the strict sense, it constitutes a complete characteristic surface and can be used because of the conservation of the Klein-Gordon current. 

The right-moving and left-moving components of the Minkowski field operator \(\hat{\phi}_M\) can be expanded as
\begin{align}
    \hat{\phi}_{M}(U_M) &= \int_0^\infty \frac{d\omega_0}{\sqrt{4\pi \omega_0}} \left[ \overrightarrow{\hat{a}}_0(\omega_0)\, e^{-i \omega_0 U_M} + \overrightarrow{\hat{a}}_0^\dagger(\omega_0)\, e^{i \omega_0 U_M} \right], \label{Apn2.1} \\
    \hat{\phi}_{M}(V_M) &= \int_0^\infty \frac{d\omega_0}{\sqrt{4\pi \omega_0}} \left[ \overleftarrow{\hat{b}}_0(\omega_0)\, e^{-i \omega_0 V_M} + \overleftarrow{\hat{b}}_0^\dagger(\omega_0)\, e^{i \omega_0 V_M} \right]. \label{Apn2.2}
\end{align}

Similarly, the right-moving and left-moving components of the Rindler field operator \(\hat{\phi}_1\) can be expanded as
\begin{align}
    \hat{\phi}_{1}(u_1) &= \int_0^\infty \frac{d\omega}{\sqrt{4\pi \omega}} \left[ \overrightarrow{\hat{a}}_1(\omega)\, e^{-i \omega u_1} + \overrightarrow{\hat{a}}_1^\dagger(\omega)\, e^{i \omega u_1} \right], \label{Apn2.3} \\
    \hat{\phi}_{1}(v_1) &= \int_0^\infty \frac{d\omega}{\sqrt{4\pi \omega}} \left[ \overleftarrow{\hat{b}}_1(\omega)\, e^{-i \omega v_1} + \overleftarrow{\hat{b}}_1^\dagger(\omega)\, e^{i \omega v_1} \right]. \label{Apn2.4}
\end{align}

Starting from Eq.~(\ref{Apn1.3}), we multiply both sides by \(\int_{-\infty}^\infty \frac{d u_1}{\sqrt{2\pi}} e^{i \omega' u_1}\), resulting in
\begin{equation}
  \begin{split}
       \int_{-\infty}^\infty \frac{du_1}{\sqrt{2\pi}} e^{i\omega' u_{1}}\;\hat{\phi}_1 
       =  \int_{0}^\infty \frac{d\omega}{\sqrt{2\pi}}\frac{1}{\sqrt{2\omega}} \\
       \int_{-\infty}^\infty \frac{du_1}{\sqrt{2\pi}}\;\bigg(\overrightarrow{\hat{a}}_1(\omega)\; e^{i(\omega'-\omega) u_{1}}+\overrightarrow{\hat{a}}_1^\dagger(\omega)\; e^{i(\omega+\omega')u_{1}}\bigg) ,
       \label{Apn2.5}
  \end{split}
\end{equation}

For $\omega' > 0$, only the first term contributes due to orthogonality, yielding
\begin{equation}
   \int_{-\infty}^\infty \frac{d u_1}{\sqrt{2\pi}} e^{i \omega' u_1} \hat{\phi}_1
   = \frac{\overrightarrow{\hat{a}}_1(\omega')}{\sqrt{2 \omega'}}.
   \label{Apn2.6}
\end{equation}

We now compute the same integral using the Minkowski field operator $\hat{\phi}_M$ from Eq.~(\ref{Apn2.1}),
\begin{equation}
\begin{split}
\int_{-\infty}^\infty \frac{du_1}{\sqrt{2\pi}} e^{i\omega' u_{1}}\hat{\phi}_{M} 
=  \frac{1}{2\pi} \int_{0}^\infty \frac{d\omega_0}{\sqrt{2\omega_0}} \int_{-\infty}^\infty du_1 \\
\bigg(\overrightarrow{\hat{a}}_0(\omega_0)\;e^{-i\omega_0 U_M} e^{i\omega'u_1}
+\overrightarrow{\hat{a}_0^\dagger}(\omega_0)\;e^{i\omega_0 U_M}e^{i\omega'u_1}\bigg).
\end{split}
\label{Apn2.7}
\end{equation}

Setting $\omega' = \omega > 0$, and using the coordinate transformation $U_M = -\frac{1}{a} e^{-a u_1}$, we identify the Bogoliubov coefficients as
\begin{align}
   \overrightarrow{\alpha_{10}^*}(\omega, \omega_0) &= \sqrt{\frac{\omega}{\omega_0}} \int_{-\infty}^\infty \frac{d u_1}{2\pi} 
   e^{i\omega u_1} \exp\left(i \frac{\omega_0}{a} e^{-a u_1}\right), \label{Apn2.8} \\
   \overrightarrow{\beta_{10}^*}(\omega, \omega_0) &= -\sqrt{\frac{\omega}{\omega_0}} \int_{-\infty}^\infty \frac{d u_1}{2\pi} 
   e^{i\omega u_1} \exp\left(-i \frac{\omega_0}{a} e^{-a u_1}\right). \label{Apn2.9}
\end{align}

After evaluation, these integrals yield
\begin{align}
 \overrightarrow{\alpha_{10}^*}(\omega, \omega_0) &= \frac{1}{2 \pi a} \sqrt{\frac{\omega}{\omega_0}} \; \Gamma\left(-\frac{i \omega}{a}\right) \left( \frac{\omega_0}{a} \right)^{\frac{i \omega}{a}} e^{\frac{\pi \omega}{2a}}, \label{Apn2.10}\\
 \overrightarrow{\beta_{10}^*}(\omega, \omega_0) &= -\frac{1}{2 \pi a} \sqrt{\frac{\omega}{\omega_0}}\; \Gamma\left(-\frac{i \omega}{a}\right)\;\left( \frac{\omega_0}{a} \right)^{\frac{i \omega}{a}} e^{-\frac{\pi \omega}{2a}} . \label{Apn2.11}
\end{align}

\section{Derivation of Sec.~VI\label{Apn3}}

We derive the general solution of Eq.~(\ref{Eq:6.1.0.1}), which governs flux-preserving conformal transformations,
\begin{equation}
\frac{1}{\left( \frac{\partial u}{\partial U} \right)^2}
\left[
\left(\frac{\partial u}{\partial U}\right)
\left(\frac{\partial^3 u}{\partial U^3}\right)
-\frac{3}{2}
\left(\frac{\partial^2 u}{\partial U^2}\right)^2
\right]
-\alpha^2
\left(\frac{\partial u}{\partial U}\right)^2
=0, \label{Apn3.1}
\end{equation}

Equation~(\ref{Apn3.1}) is equivalent to a Schwarzian differential equation. Here we derive its general solution using a standard method.

Multiplying Eq.~(\ref{Apn3.1}) by
$2(\partial u/\partial U)^2$, we obtain
\begin{equation}
2u'u'''-3(u'')^2  = 2\alpha^2(u')^4,\label{Apn3.2}
\end{equation}
where primes denote differentiation with respect to $U$.

Comparing Eq.~(\ref{Apn3.2}) with the standard nonlinear equation
\begin{equation}
2y'y'''-3(y'')^2
=
f(x)(y')^2
+
g(x)(y')^4, \label{Apn3.3}
\end{equation}
we identify
\begin{equation}
f(x)=0,
\qquad
g(x)=2\alpha^2, \label{Apn3.4}
\end{equation}

Following the standard solution procedure described in Ref.~\cite{book}, one introduces auxiliary functions $u(y)$ and $w(x)$, which satisfy
\begin{equation}
u(y)
=
Ae^{\frac{\alpha}{\sqrt2}y}
+
Be^{-\frac{\alpha}{\sqrt2}y}, \label{Apn3.5}
\end{equation}
and
\begin{equation}
w(x)=Cx+D,\label{Apn3.6}
\end{equation}
where $A,B,C,$and  $D$ are integration constants.

These functions are related through
\begin{equation}
\int\frac{dy}{u^2(y)}
=
\int\frac{dx}{w^2(x)}
+\text{const}, \label{Apn3.7}
\end{equation}
Evaluating the integrals and absorbing the integration constants into new parameters yields
\begin{equation}
u(U)
=
\frac{1}{\alpha}
\ln\left|
\frac{aU+b}{cU+d}
\right|,\label{Apn3.8}
\end{equation}
where $a,b,c,$ and $d$ are arbitrary constants satisfying

$ad-bc\neq0$

The same method can be applied to Eq.~(\ref{Eq:7.0.0.1}),
\begin{equation}
\{u_1,u_2\}
-\frac{\alpha^2}{2}
\left(\frac{\partial u_1}{\partial u_2}\right)^2
=
-\frac{\alpha^2}{2}, \label{Apn3.9}
\end{equation}
which differs from Eq.~(\ref{Apn3.1}) only by the presence of the constant term on the right-hand side. Rewriting it as
\begin{equation}
\begin{split}
\frac{1}{\left(\frac{\partial u_1}{\partial u_2}\right)^2}
\left[
\left(\frac{\partial u_1}{\partial u_2}\right)
\left(\frac{\partial^3u_1}{\partial u_2^3}\right)
-\frac{3}{2}
\left(\frac{\partial^2u_1}{\partial u_2^2}\right)^2
\right]
-\frac{\alpha^2}{2}
\left(\frac{\partial u_1}{\partial u_2}\right)^2\\
= -\frac{\alpha^2}{2}, \label{Apn3.10}
\end{split}
\end{equation}
and comparing with the standard equation
\begin{equation}
2y'y'''-3(y'')^2
=
f(x)(y')^2
+
g(x)(y')^4, \label{Apn3.11}
\end{equation}
We identify,
\begin{equation}
f(x)=-\alpha^2,  \qquad g(x)=\alpha^2,  \label{Apn3.12}
\end{equation}
Following the same procedure as above (see Ref.~\cite{book}), the general solution is found to be
\begin{equation}
e^{\alpha u_1} =
=
\frac{A e^{\alpha u_2}+B}
{C e^{\alpha u_2}+D},
\qquad
AD-BC\neq0,  \label{Apn3.13}
\end{equation}
where $A,B,C,$  and $D$ are arbitrary constants. An identical derivation yields the corresponding solution for the $v$-sector.
\begin{acknowledgments}
The authors thank the BITS Pilani Hyderabad campus for providing the necessary infrastructure for this research.
\end{acknowledgments}
\bibliography{Ref.bib}
\end{document}